\documentclass[conference]{IEEEtran}
\IEEEoverridecommandlockouts
\usepackage{cite}
\usepackage{amsmath,amssymb,amsfonts}
\usepackage[dvips]{graphicx}
\usepackage{textcomp}
\usepackage{xcolor}

\usepackage{mathtools}
\usepackage{subcaption} 
\usepackage{multirow}
\usepackage{tikz}
\usepackage{pgfplots}
\usepackage{url}
\usepackage{balance}
\pgfplotsset{compat=1.14}

\usepackage[normalem]{ulem}
\usepackage{balance}
\usepackage{enumitem}
\usepackage{soul}
\usepackage{tikz}

\usepackage{algpseudocode}
\usepackage{algorithm}

\usepackage{caption} 
\usepackage{textcomp}
\newtheorem{analysis-rule}{Rule}[subsection]
\newtheorem{analysis-sub-rule}{Rule}[analysis-rule]

\usepackage{listings}
\usepackage{color}
\usepackage{cite}
\usepackage{amssymb}

\DeclareMathAlphabet{\mathpzc}{OT1}{pzc}{m}{it}

\definecolor{dkgreen}{rgb}{0,0.6,0}
\definecolor{gray}{rgb}{0.5,0.5,0.5}
\definecolor{mauve}{rgb}{0.58,0,0.82}

\newcommand{\oursys}{\textsc{AARSynth}}

\newcommand\redsout{\bgroup\markoverwith{\textcolor{red}{\rule[0.5ex]{2pt}{0.4pt}}}\ULon}

\definecolor{fgreen}{rgb}{0.0, 0.5, 0.0}

\newcommand{\stitle}[1]{\vspace{1ex}\noindent\textup{\textbf{#1}}}
\newcommand{\myNum}[1]{(\emph{#1})}
\newcommand{\CASE}[1]{\STATE \textbf{case} #1\textbf{:} \begin{ALC@g}}
\newcommand{\ENDCASE}{\end{ALC@g}}

\newcommand{\DEFAULT}{\STATE \textbf{default:} \begin{ALC@g}}
\newcommand{\ENDDEFAULT}{\end{ALC@g}}
\newcommand{\DEFAULTLINE}[1]{\STATE \textbf{default:} }

\newcommand{\realR} {\mathbb{R} }
\newcommand{\bigB} {\mathcal{B} }
\newcommand{\bigC} {\mathcal{C} }
\newcommand{\bigH} {\mathcal{H} }
\newcommand{\bigL} {\mathcal{L} }
\newcommand{\bign} {N}

\newcommand{\bigR} {\mathcal{R} }
\newcommand{\bigS} {\mathcal{S} }
\newcommand{\bigx} {\mathcal{X} }
\newcommand{\bigy} {\mathcal{Y} }

\newcommand{\atta} {\mathpzc{\textbf{a}}}
\newcommand{\smallb} {b}
\newcommand{\smallc} {\mathpzc{c} }
\newcommand{\smalld} {\mathpzc{d} }
\newcommand{\smalle} {\mathpzc{e} }

\newcommand{\smallg} {g}
\newcommand{\smallh} {\mathpzc{\textbf{h}}}
\newcommand{\smalli} {\mathpzc{i} }
\newcommand{\smallj} {j}
\newcommand{\smallk} {k}
\newcommand{\smallm} {\mathpzc{m} }
\newcommand{\smalln} {\mathpzc{n} }

\newcommand{\smallr} {\mathpzc{r} }
\newcommand{\smalls} {\mathpzc{s} }

\newcommand{\smallu} {\mathpzc{u} }
\newcommand{\attv} {\mathpzc{\textbf{v}}}
\newcommand{\smallw} {\mathpzc{w} }
\newcommand{\smallx} {\mathpzc{x} }
\newcommand{\smally} {\mathpzc{y} }
\newcommand{\smallz} {\mathpzc{\textbf{z}} }

\newcommand{\eos} {\mathpzc{eos} }
\newcommand{\sos} {\mathpzc{sos} }
\newcommand{\encoder} {\text{encoder} }
\newcommand{\decoder} {\text{decoder} }
\newcommand{\softmax} {\text{softmax} }
\newcommand{\emb} {\text{emb} }
\newcommand{\col} {\mathpzc{col} }

\usepackage{comment}

\newcommand{\one} {\mathpzc{1} }
\newcommand{\two} {\mathpzc{2} }
\usepackage{multirow}
\usepackage{array}

\def\BibTeX{{\rm B\kern-.05em{\sc i\kern-.025em b}\kern-.08em
    T\kern-.1667em\lower.7ex\hbox{E}\kern-.125emX}}
\begin{document}

\IEEEoverridecommandlockouts
 \IEEEpubid{\makebox[\columnwidth]{978-1-7281-6251-5/20/\$31.00 \copyright 2020 IEEE} 
\hspace{\columnsep}\makebox[\columnwidth]{ }}

\title{App-Aware Response Synthesis for User Reviews}

\author{\IEEEauthorblockN{Umar Farooq\textsuperscript{\textasteriskcentered}\thanks{\textsuperscript{\textasteriskcentered} Equal contribution.}, A.B. Siddique\textsuperscript{\textasteriskcentered}, Fuad Jamour, Zhijia Zhao, Vagelis Hristidis}
\IEEEauthorblockA{University of California, Riverside\\
ufaro001@ucr.edu, msidd005@ucr.edu, fuadj@ucr.edu, zhijia@cs.ucr.edu, vagelis@cs.ucr.edu
}}

\maketitle

\begin{abstract}
Hundreds of thousands of mobile app users post their reviews online.
Responding to user reviews promptly and satisfactorily improves application ratings, which is key to application popularity and success.
The proliferation of such reviews makes it virtually impossible for developers to keep up with responding manually.
To address this challenge, recent work has shown the possibility of automatic response generation by training a seq2seq model with a large collection of review-response pairs. However, because the training review-response pairs are aggregated from many different apps, it remains challenging for such models to generate \emph{app-specific} responses, which, on the other hand, are often desirable as apps have different features and concerns.
Solving the challenge by simply building an app-specific generative model per app (i.e., training the model with review-response pairs of a single app) may be insufficient because individual apps have limited review-response pairs, and such pairs typically lack the relevant information needed to respond to a new review.

To enable \emph{app-specific} response generation, this work proposes {\oursys}: an app-aware response synthesis system. The key idea behind {\oursys} is to augment the seq2seq model with information specific to a given app. Given a new user review, {\oursys} first retrieves the top-K most relevant app reviews and the most relevant snippet from the app description. The retrieved information and the new user review are then fed into a fused machine learning model that integrates the seq2seq model with a machine reading comprehension model. The latter helps digest the retrieved reviews and app description. Finally, the fused model generates a response that is customized to the given app. We evaluated {\oursys} using a large corpus of reviews and responses from Google Play. The results show that {\oursys} outperforms the state-of-the-art system by $22.2\%$ on BLEU-4 score. Furthermore, our human study shows that {\oursys} produces a statistically significant improvement in response quality compared to the state-of-the-art system. 

\end{abstract}

\begin{IEEEkeywords}
App reviews, Natural language generation, Machine translation.
\end{IEEEkeywords}

\section{Introduction}
The wide adoption of smartphones has created a large and growing market for mobile apps.
Recent studies~\cite{statista} predicted that the number of smartphone users will reach $3.8$ billion worldwide by $2021$, projecting a market of trillion dollars for mobile apps by $2023$.
These apps are typically distributed through app stores such as Apple App Store and Google Play.
App stores allow users to give their feedback, ask questions, and publicly express their levels of satisfaction with an app through reviews and ratings: positive reviews and ratings are important factors to acquire and retain users.
App developers can respond to user feedback to maintain and improve their app reviews and ratings. According to Google Play, responding to user reviews leads to an increase of $0.7$ stars for an app on average~\cite{googleplayresponse}.
While developers recognize the importance of responding to user reviews promptly, the proliferation of reviews makes it virtually impossible to manually provide responses to all the reviews.

\begin{figure}[t!]
\centering
\includegraphics[width=\linewidth]{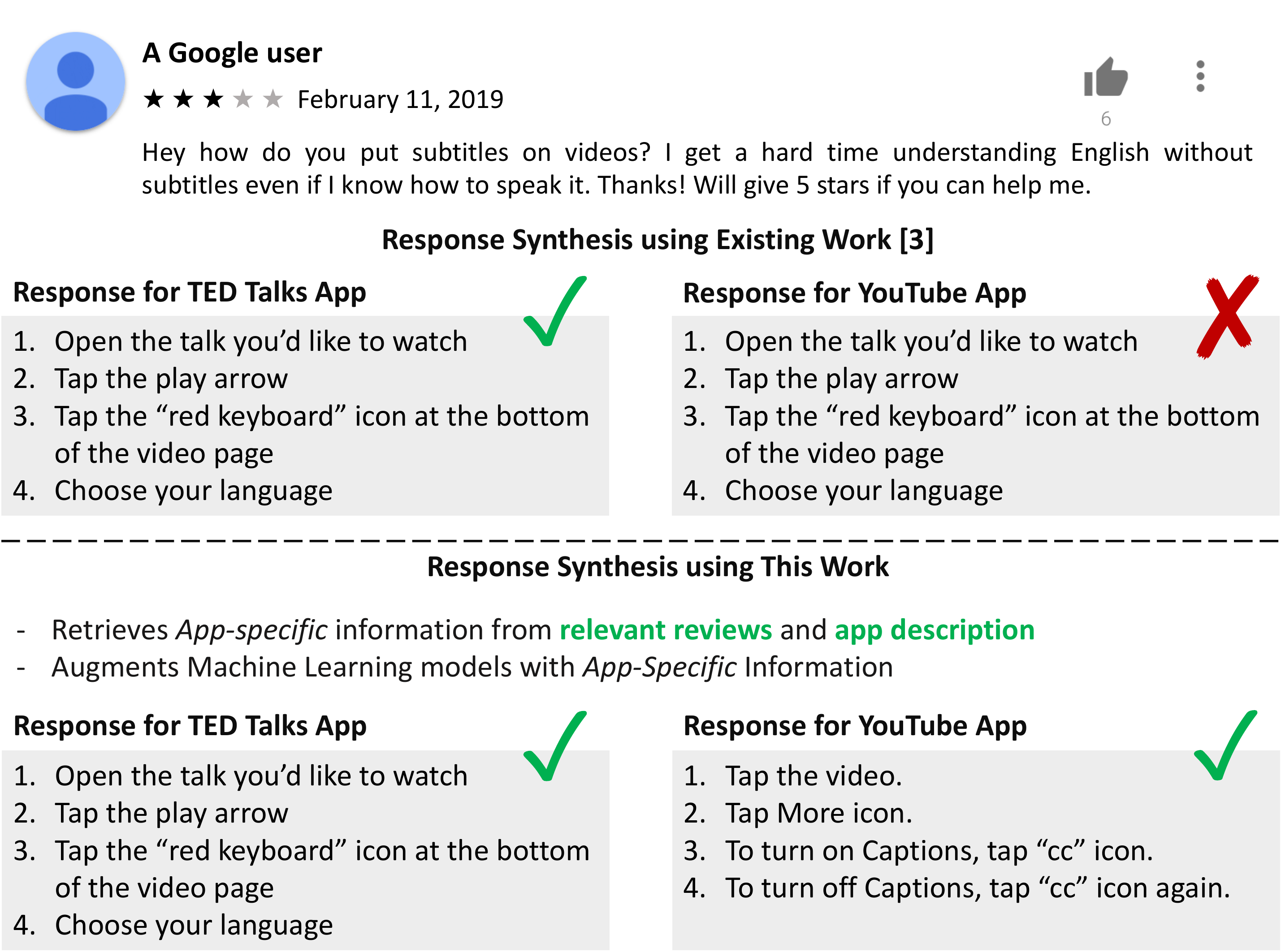}
\caption{Example user review and automatic responses: The response generated by RRGen~\cite{rrgen} is suitable for TED Talks but not for YouTube. By contrast, our system generates responses \emph{specific} to the considered app with the help of the most relevant reviews and the app description.}
\label{fig:intro}
\end{figure}

\begin{figure*}[t!]
\centering
\includegraphics[width=0.9\textwidth]{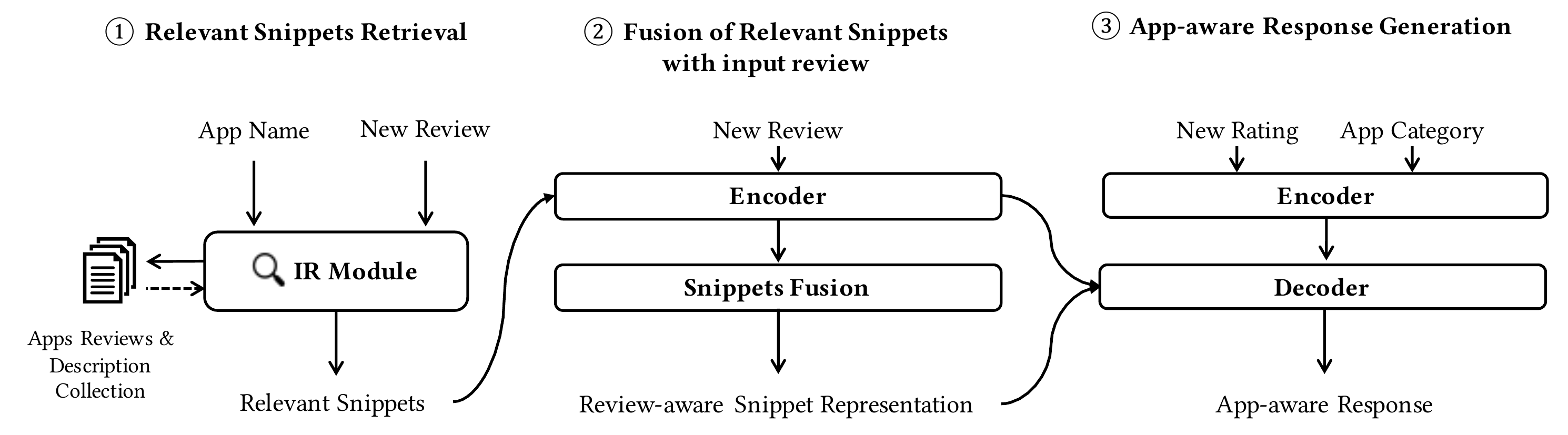}
\vspace{-5pt}
\caption{Overview of {\oursys}.} 
\label{fig:intro-example}
\vspace{-10pt}
\end{figure*}

Given the importance of responding to user reviews and the impossibility of manually responding to each review, it has become crucial to automatically synthesize responses. 
Little research has been done to build machine learning models for automatically synthesizing responses to app reviews and questions. 
RRGen~\cite{rrgen} stands out as the first effort in this direction, where the authors proposed using an attention-based sequence-to-sequence (a.k.a. seq2seq) neural model~\cite{sutskever2014sequence}. The authors trained their model using a dataset of review-response pairs for a large collection of apps hosted on Google Play.
RRGen generates satisfactory responses to reviews common among many apps (e.g., ``{\em lots of ads}''); however, it fails to synthesize responses specific to an app -- see Figure~\ref{fig:intro} for an example where RRGen generates the same response for two different apps (TED Talks and YouTube).
In this example, RRGen was able to generate an appropriate response for TED Talks but not for YouTube because its training data happened to have information specific to TED Talks.

There are two key challenges in generating app-specific responses effectively. First, there may be a lack of sufficient review-response pairs for the given app to reliably train a generative model. Second, appropriate responses may not be available in any of the existing training pairs.
To address the first challenge, we leverage the review-response pairs from other apps.
For the second challenge, we propose to feed the generative model with the most relevant snippets extracted from the app's description and its existing user reviews -- important sources of information that have been neglected or under-utilized by prior work.
In fact, our analysis of $10$ trending apps on Google Play revealed that $50$\% of the analyzed apps provide FAQ sections and $32.3$\%  have reviews with snippets closely relevant to answer app-specific questions raised in other reviews of the same app.

Figure~\ref{fig:intro-example} shows an overview of {\oursys}. Given a new user review for an app, our information retrieval (IR) module first retrieves relevant text snippets from the app's description and existing user reviews.
Then, the snippets fusion layer uses these snippets to build a review-aware representation (i.e., a representation of the snippets that is associated with the input review) similarly to machine reading comprehension models~\cite{wang2017r,seo2016bidirectional} (MRC).
Finally, the review-aware snippets representation and the input review representation along with other app-specific features produced by our encoder are passed to the decoder to produce an app-specific response.

Note that neither an MRC model nor a seq2seq model by themselves are sufficient for the above task.
On one hand, an MRC model typically produces a span text -- a substring extracted from a textual document, which is hard to comprehend as the response to a review. 
On the other hand, despite that a seq2seq model can produce free-form text, it can not use the relevant snippets unless they are transformed into a representation that is aware of the input review.
Our fusion between the seq2seq and MRC models resolves the limitations of each model in the context of app response generation.

We  evaluated {\oursys} using a large corpus of reviews, responses, and app descriptions collected from Google  Play.
Our results show that {\oursys} outperforms the state-of-the-art system RRGen~\cite{rrgen} by $22.2$\% in BLUE-4 score -- a widely used metric for text generation.
Moreover, our human study using Amazon Mechanical Turk shows that the responses generated by {\oursys} better address user concerns, are app-specific, and are more fluent than the responses generated by RRGen with a statistically significant difference.

In summary, this paper makes the following contributions:
\begin{itemize}[leftmargin=6mm]
\item
It proposes a novel neural architecture that fuses seq2seq and machine reading comprehension models to synthesize free-form responses specific to an app.
\item
It releases a large dataset\footnote{Available at \url{https://github.com/AARSynth/Dataset}}
that consists of more than $570$K review-response pairs and more than $2$ million user reviews for $103$ popular applications.
\item
It conducts extensive experimental analysis using our large dataset and compares {\oursys} against the state-of-the-art systems. The evaluation using automatic metrics and real-user studies confirms the competitiveness of {\oursys} with a statistically significant improvements.
\end{itemize}

In the following, we first present a brief review on attentional encoder-decoder models and information retrieval techniques in Section~\ref{sec:background}. Then, we present {\oursys} in Section~\ref{sec:model}, followed by our experimental setup (Section~\ref{sec:experiments}) and evaluation (Section~\ref{sec:evals}). Finally, we discuss the related work in Sections~\ref{sec:related} and conclude this work in Section~\ref{sec:conclusion}.

\section{Background}
\label{sec:background}

{\oursys} is built on top of seq2seq neural networks and state-of-the-art information retrieval techniques. This section offers a brief introduction to these topics.

\subsection{Attentional Encoder-Decoder Model}
\label{sec:seq2seq}
\begin{figure}[t!]
\centering
\includegraphics[width=\linewidth]{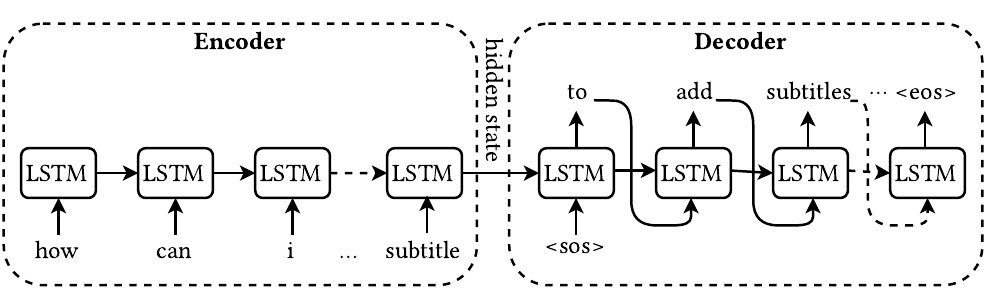}
\caption{Overview of the seq2seq neural model.}
\label{fig:seq2seq}
\end{figure}

The goal of an encoder-decoder model, like seq2seq~\cite{sutskever2014sequence}, is to synthesize a target sequence (e.g., developer response)
$\bigy = (\smally_\one, \smally_\two, ...,\smally_\smallm)$
given an input sequence (e.g., user review)
$\bigx = (\smallx_\one, \smallx_\two, ..., \smallx_\smalln)$, where $\smallm$ and $\smalln$ are target and input sequence lengths.
Figure~\ref{fig:seq2seq} presents a high level overview of the seq2seq model that shows the encoder and the decoder parts.

\stitle{Encoding.}
The encoder reads an input sequence of length $\smalln$, one token at a time until it encounters the end of sequence token (i.e., $<\eos>$). It transforms the sequence into hidden states $\bigH = (\smallh_\one, \smallh_\two,...,\smallh_\smalln)$ by applying a Recurrent Neural Network (RNN), such as Long Short-Term Memory (LSTM)~\cite{hochreiter1997lstm}. 
Specifically, it transforms input token $\smallx_\smalli$ to hidden state $\smallh_\smalli = \encoder(\smallh_{\smalli-\one}, \emb(\smallx_\smalli))$, where $\encoder(\cdot)$ is a non-linear mapping function, $\emb(\smallx_\smalli)$ is the word embedding of input token $\smallx_\smalli$, and $\smallh_{\smalli-\one}$ is the previous hidden state.

\begin{figure}[t!]
\centering
\includegraphics[width=\linewidth]{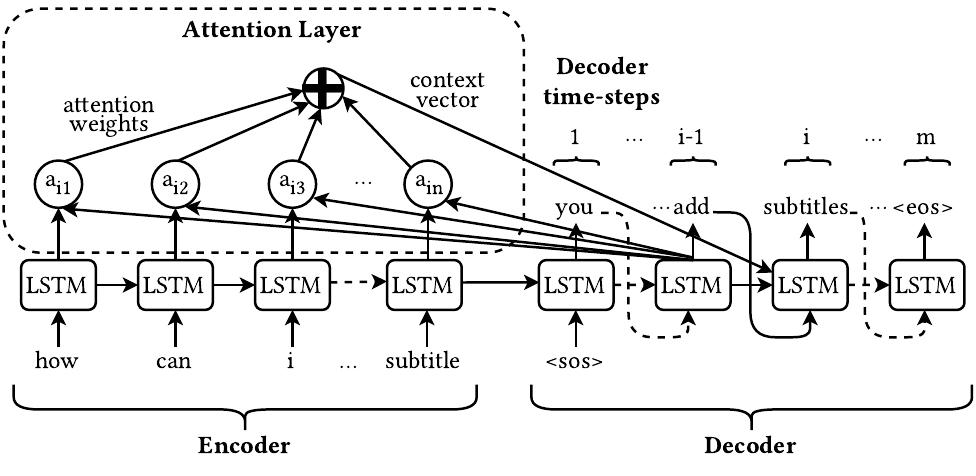}
\caption{Attention-based seq2seq neural model.}
\label{fig:seq2seq-attention}
\end{figure}

\stitle{Decoding.}
The decoder is initialized with the encoder's last hidden state $\smallh_\smalln$ and the start of sequence token (i.e., $<\sos>$), then it utilizes another RNN to generate the target sequence $\hat \bigy$.
The decoder also generates one token at a time, until the end token (i.e., $<\eos>$) is generated.
The generation, at time-step $\smalli$, is conditioned on the previously generated words $\hat \smally_{\smalli-\one}$,..., $\hat \smally_\one$, and the decoder's current hidden state $\smallh^\prime_\smalli$, according to the following probability distribution:
\begin{equation*}
  P(\hat \smally_\smalli | \hat \smally_{\smalli-\one},...,\hat \smally_\one, \bigx)= \softmax(\decoder(\smallh^\prime_\smalli, \hat \smally_{\smalli-\one})),
  \label{eq:enc-dec}
\end{equation*}
where $\decoder(\cdot)$ is a non-linear mapping function and $\softmax(\cdot)$ converts the given vector into a  probability distribution.
The encoder-decoder model is trained jointly by minimizing the negative log-likelihood loss of the given $\bign$ training input-target sequence pairs of the form ($\bigx^\smalli$, $\bigy^\smalli$):
\begin{equation*}
\label{eq:s2s_loss}
 \bigL(\theta)= - \min_\theta \frac{\one}{\bign} \sum_{\smalli=\one}^{
 \bign} \log p_\theta (\bigy^\smalli | \bigx^\smalli),
\end{equation*}
where $\theta$ is a set of trainable parameters estimated using optimization algorithms such as stochastic gradient descent.

\stitle{Attentional Decoding.}
Attention mechanisms~\cite{vaswani2017attention} are used in seq2seq models to pay attention to more relevant input tokens while decoding.
Figure~\ref{fig:seq2seq-attention} shows the computation of an attention vector at time-step~$\smalli$. 
While decoding at time-step~$\smalli$, the $\decoder(\cdot)$ is not only conditioned on the decoder's current hidden state $\smallh^\prime_\smalli$ and the previous generations $\smally_{\smalli-\one}$,..., $\smally_\one$, but also on the attention context vector $\attv_\smalli$, which is computed as:
\begin{equation*}
\label{eq:context}
 \attv_\smalli = \sum_{\smallk=\one}^{\smalln} \atta_{\smalli \smallk} \smallh_\smallk,
\end{equation*}
where $\atta_{\smalli \smallk}$ is the attention weight for hidden state $\smallh_\smallk$,  capturing how relevant is encoder's $\smallk$-th hidden state for predicting token $\hat \smally_\smalli$ while considering the decoder's previous hidden state $\smallh^\prime_{\smalli-\one}$. The attention weight $\atta_{\smalli \smallk}$ at time-step~$\smalli$ can be computed as:
\begin{equation*}
\label{eq:attention}
 \atta_{\smalli \smallk} = \frac{\exp(\smalle_{\smalli \smallk})}
 {\sum_{\smallj=\one}^{\smalln} exp(\smalle_{\smalli \smallj})},
\end{equation*}
where 
 $\smalle_{\smalli \smallk} =  \text{align}(\smallh^\prime_{\smalli-\one}, \smallh_\smallk)$
and \text{align($\cdot$)} is an alignment model implemented as a Multi-Layer Perception (MLP) unit. 
The attentional seq2seq neural network is trained jointly.
We employ the seq2seq framework as a basic building block in our proposed approach for response synthesis.

\subsection{Relevant Document Retrieval}
\label{sec:info-ret}

\begin{figure}[t!]
\centering
\includegraphics[width=\linewidth]{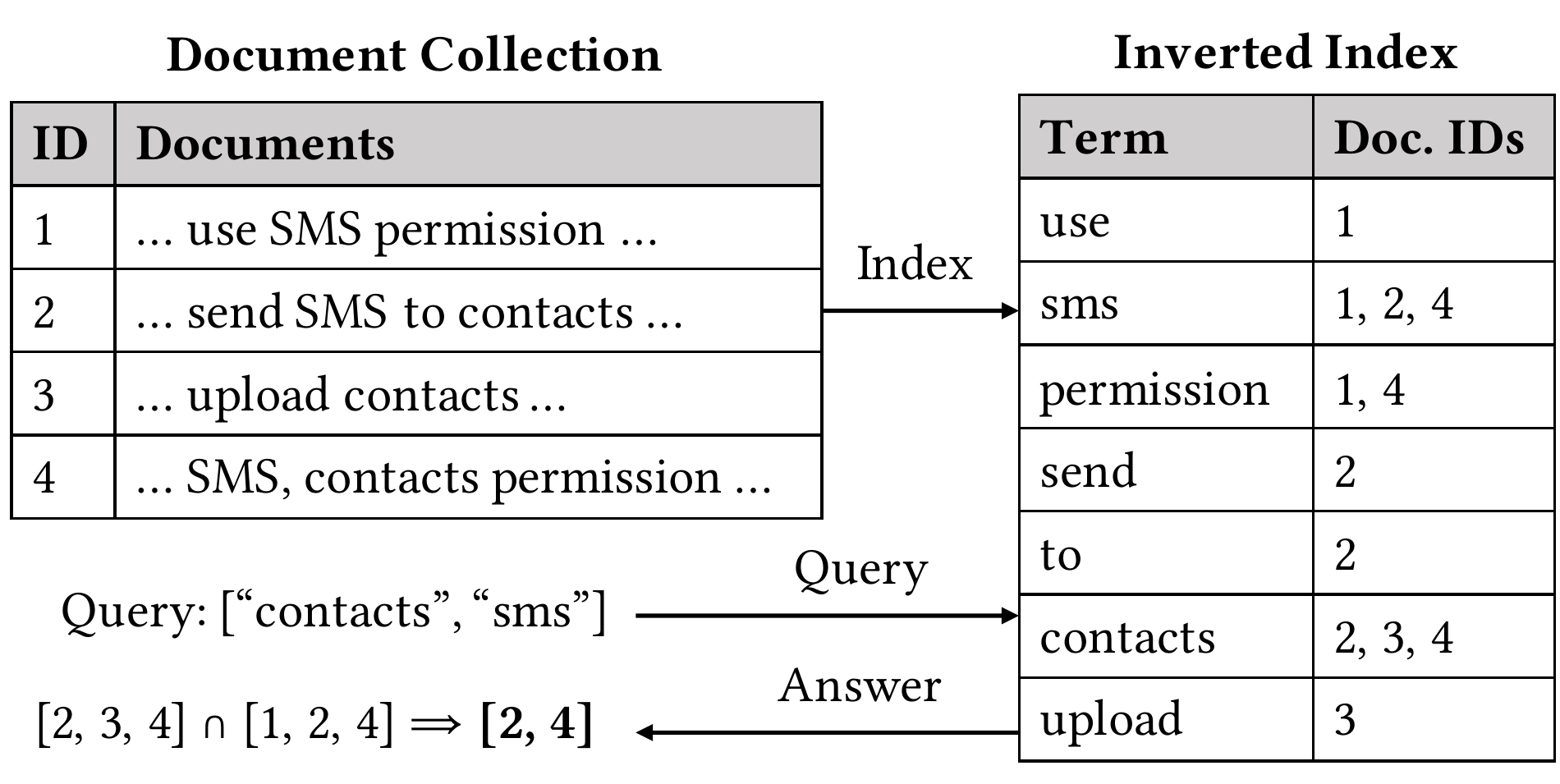}
\caption{Overview of IR indexing and searching.}
\label{fig:ir-background}
\end{figure}

Given a collection of documents (e.g., app description and reviews) and a query (e.g., new review), an IR system returns a subset of documents relevant to the query.
Figure~\ref{fig:ir-background} illustrates how an IR system works: the system creates an index for the given document collection, based on which it finds relevant document(s) for a given query.

\stitle{Document Indexing.}
IR systems create inverted indexes to facilitate faster document retrieval.
An inverted index consists of a set postings lists; a postings list is a list of individual postings, each of which provides information about occurrences of a term (i.e., word) in the document collection including document id and the number of occurrences of the term in each document that contains the term -- term frequency (TF).
We use Apache Lucene~\cite{lucene}, which uses SkipList~\cite{pugh1990skip} to implement postings lists for fast retrieval.  
 
\stitle{Searching and Ranking.}
IR systems typically use vector space retrieval with Term Frequency-Inverse Document Frequency (TF-IDF) weighting~\cite{baeza1999modern}. 
TF captures the importance of a term in a document, and IDF captures the significance of the term in the whole collection. 
The BM25~\cite{bm25} algorithm is widely used by IR systems to improve search engine relevance. BM25 scores a document $D$ for input query $Q$ with terms $q_1,q_2,...,q_n$ as follows: 

\begin{equation*} 
\label{eq:bm25} 
\text{score}(D,Q) = \sum_{i=1}^{n} \text{IDF}(q_i) \cdot \frac{f(q_i, D) \cdot (k_1 + 1)}{f(q_i, D) + k_1 \cdot (1 - b + b \cdot \frac{|D|}{\text{avgdl}})},
\end{equation*}
where $f(q_i, D)$ represents the term frequency of $q_i$ in the document $D$, the document $D$ with length $|D|$, and $avgdl$ is the average document length in the whole collection, $k_1 \in [1.2, 2.0]$ and $b =0.75$.
In the context of {\oursys}, we use IR to retrieve the most relevant text snippets from description and reviews of the given app.

\section{App-Aware Response Synthesis}
\label{sec:model}

\begin{figure}[t!]
\centering
\includegraphics[width=\linewidth]{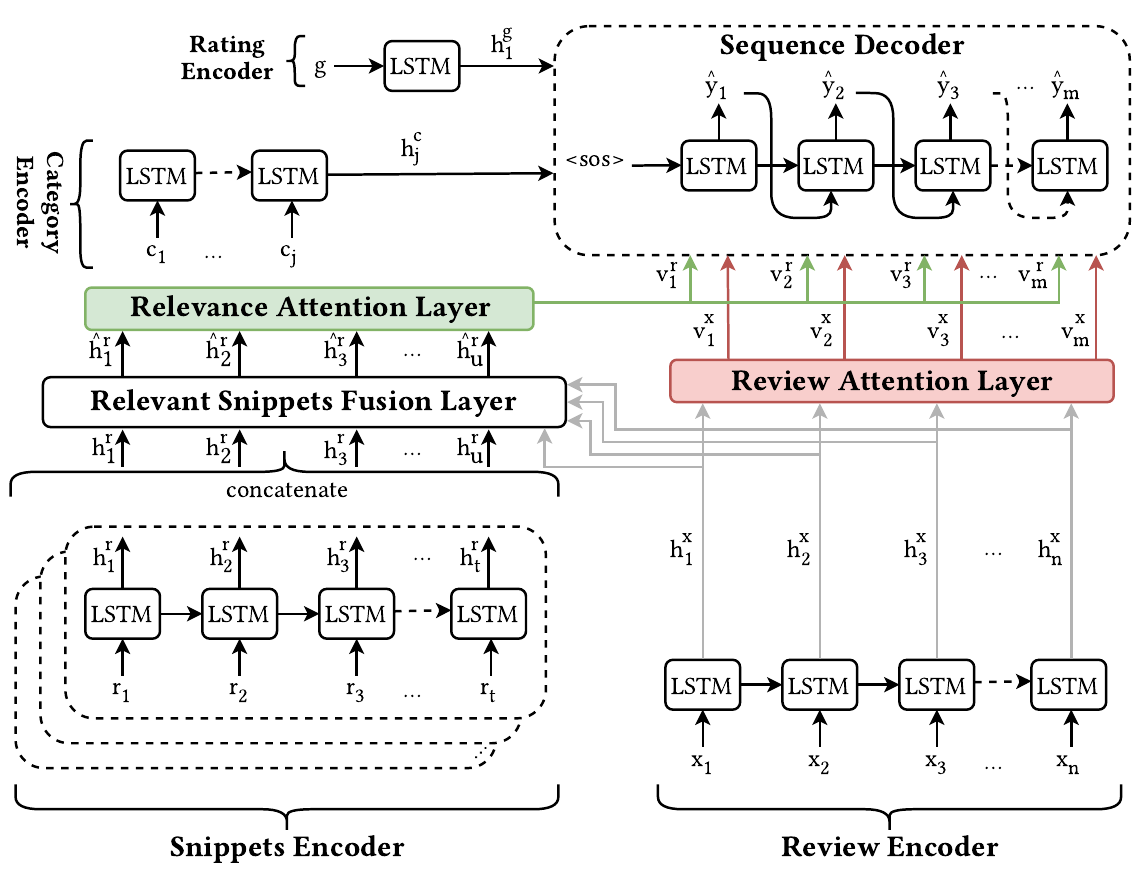}
\caption{Architecture of {\oursys}'s neural model.}
\label{fig:seq2seq-model}
\end{figure}

In this section, we present {\oursys} (see Figure~\ref{fig:intro-example} for an overview). In the following, we first introduce the IR module, then provide an overview of the neural model followed by the details of each of its major components.

\subsection{IR Module}
\label{sec:ir}

{\oursys} employs IR module as follows. First, it builds an inverted index for all the app descriptions and reviews. 
Then, it takes a new user review and the app name as inputs, and retrieves top-k relevant documents (i.e., reviews and description of the given app). To get the most relevant text snippets from app description and reviews, we leverage Lucene~\cite{lucene} Highlighter and Sentence Detector~\cite{manning2014stanford} to extract the most relevant snippets from the retrieved document(s). We refer to these text snippets as {\em relevant snippets}.

\subsection{Overview of Neural Model}
\label{sec:neural}
Figure~\ref{fig:seq2seq-model} shows the architecture of the proposed app-aware response synthesis model, which consists of the following components:
\myNum{i} input encoding layers to transform textual inputs to high dimensional contextual representations;
\myNum{ii} relevant snippets fusion layer to produce a review-aware representation of the relevant snippets;
\myNum{iii} attention layers to compute attention weights that capture the significance of tokens in the input review and its relevant snippets; and
\myNum{iv} sequence decoding layer that fuses all the information from the previous layers to ultimately synthesize an app-aware response.
We next explain each component in detail.

\subsection{Input Encoding Layers}
\label{sec:input}

\stitle{Review Encoder.}
To encode an input review $\bigx = (\smallx_\one, \smallx_\two, ..., \smallx_\smalln)$, the review encoder first maps each word $\smallx_\smalli$ to a high dimensional vector space (i.e., word embedding) $\emb^\smallx_\smalli$, then an RNN is utilized to produce a new $\smalld$-dimensional representation $\bigH^\bigx = (\smallh^\smallx_\one, \smallh^\smallx_\two,...,\smallh^\smallx_\smalln) \in \realR^{\smalld \times \smalln}$ of all the tokens in the input review, where LSTM is used as an RNN. The token encoding at time-step $\smalli$ is computed using an LSTM as follows:
\begin{equation*}
  \smallh^\smallx_\smalli = \text{LSTM}(\smallh^\smallx_{\smalli-\one}, \emb^\smallx_\smalli).
  \label{eq:encoder*}
\end{equation*}
This representation is used by the review attention layer to compute the attention weights vector for the sequence decoder and relevant snippets fusion layer to compute the review-aware representation of the relevant snippets.

\stitle{Snippets Encoder.}
The relevant snippets are retrieved by our IR module based on the input review and app name, and then passed to the snippets encoder, which produces a new representation $\bigH^\bigR = (\smallh^\smallr_\one, \smallh^\smallr_\two,...,\smallh^\smallr_\smallu) \in \realR^{\smalld \times \smallu}$ where $\smallu$ is the total number of tokens in all the retrieved snippets.
The representation is the result of concatenating the representations produced by the RNN for each snippet. This representation is used by the relevant snippets fusion layer to compute the importance of the words in the snippets with respect to the given user review $\bigx$, and is used by the relevance attention layer to generate an attention weights vector of the relevant snippets.

\stitle{Category and Rating Encoders.}
The category encoder produces a representation $\smallh^\smallc_\one, \smallh^\smallc_\two,...,\smallh^\smallc_\smallb$ for the category of the app and the rating encoder encodes the review rating into $\smallh^\smallg_\one \in \realR^\smalld$.
The final hidden states of these layers are passed to the sequence decoder.

\subsection{Relevant Snippets Fusion Layer}
\label{sec:mrc}
This layer associates and fuses information from the relevant snippets and the words of the input review.
First, we compute a similarity matrix $\bigS \in \realR^{\smallu \times \smalln}$ between the encodings of the snippets $\bigH^\bigR$ and the encodings of the review $\bigH^\bigx$, where $\bigS_{\smallb \smallk}$ (value at row $b$ and column $k$) represents the similarity between the $\smallb$-th word in the snippets and $\smallk$-th word in the user review, which is computed using $\bigS_{\smallb \smallk} = \alpha(\bigH^\bigR_{:\smallb}, \bigH^\bigx_{:\smallk} ) \in \realR$.
$\alpha$ is a function trained to capture the similarity between input vectors $\bigH^\bigR_{:\smallb}$ and $\bigH^\bigx_{:\smallk}$, where $\bigH^\bigR_{:\smallb}$ and $\bigH^\bigx_{:\smallk}$ are $\smallb$-th and $\smallk$-th column-vectors of $\bigH^\bigR$ and $\bigH^\bigx$, respectively.
$\alpha (\smallr, \smallx) = \smallw^\top_{(\smalls)} [ \smallr \oplus \smallx \oplus \smallr \otimes \smallu]$, where $\oplus$ is vector concatenation, $\otimes$ is element-wise multiplication, and $\smallw_{(\smalls)}$ is a trainable weight vector. Then, from the similarity matrix $\bigS$, we can get the most important snippet words with respect to the review, i.e., with the closest similarity to the user review. The attention weights for the snippet words are computed using $\smallz = \softmax(\max_{\col}(\bigS)) \in \realR^\smallu$, where $\max_{\col}$ represents $\max$ across columns. Then, the most important words in the snippets with respect to the review $\hat \bigH^\bigR \in \realR^{\smalld \times \smallu}$ can be computed by tiling the operation $\hat \smallh^\bigR$ across columns $\smallu$ times, where $\hat \smallh^\bigR = \sum_\smallb \smallz_\smallb \bigH^\bigR_{:\smallb} \in \realR^\smalld$. The matrix $\hat \bigH^\bigR \in \realR^{\smalld \times \smallu}$ represents the \emph{fused} information between user review and relevant snippets, i.e., review-aware representation of the snippets. This representation can be thought of an MRC model representation of the snippets that is fused with seq2seq representations in the sequence decoder while synthesizing the response.

\subsection{Attention Layers}
\label{sec:attention}

\stitle{Review Attention Layer.}
This layer computes the attention weights of each token in the review as follows.
It takes in the encoded representation of the review $\bigH^\bigx$ and the decoder's hidden state at previous time-step $\smallh^\prime_{\smalli-1}$, and produces an attention context vector $\attv^\smallx_\smalli$ for the decoder at time-step~$\smalli$.
The context vector $\attv^\smallx_\smalli$ captures the importance of each hidden state of the review encoder while generating token $\hat \smally_\smalli$ at time-step~$\smalli$.
The details on how to compute the context vector are presented in section~\ref{sec:seq2seq}.

\stitle{Relevance Attention Layer.}
This layer produces a representation that enables the decoder to pay more attention to the important words in the relevant snippets while generating the final response.
It takes in the review-aware representation of the relevant snippets $\hat \bigH^\bigR$ and decoder's hidden state at previous time-step $\smallh^\prime_{\smalli-1}$ and computes attention context vector $\attv^\smallr_\smalli$ for decoder at time-step~$\smalli$. The attention context vector $\attv^\smallr_\smalli$ signifies the importance of each hidden state of the review-aware representation produced by relevant snippets fusion layer (Section~\ref{sec:mrc}) that sequence decoder utilizes while synthesizing the token at time-step~$\smalli$.

\subsection{Sequence Decoder}
\label{sec:decoder}
The sequence decoder fuses the representations of the review, review rating, and category with the MRC style review-aware representation of snippets to generate the final response. The sequence decoder has a $\softmax$-based linear layer that follows RNN to map the $\smalld$-dimensional hidden states to a probability distribution over the whole vocabulary. At time-step $\smalli$, the decoder computes a conditional probability to generate $\hat \smally_\smalli$ as given below:
\begin{equation*}
\resizebox{\columnwidth}{!}{
  $P(\hat \smally_\smalli | \hat \smally_{\smalli-\one},...,\hat \smally_\one, \bigx, \bigR, \bigC, \smallg)= \decoder(\smallh^\prime_\smalli, \emb^{\hat \smally}_{\smalli-\one}, \attv^\smallx_\smalli, \attv^\smallr_\smalli, \smallh^\smallc_\smallb, \smallh^\smallg_\one).$
  }
  \label{eq:decoder}
\end{equation*}

Note that, at time-step~$\smalli$, the decoder considers its current hidden state $\smallh^\prime_\smalli$, the embedding of the token prediction $\emb^{\hat \smally}_{\smalli-\one}$ at the previous time-step, review attention vector $\attv^\smallx_\smalli$ from the review attention layer, relevant snippets attention vector $\attv^\smallr_\smalli$ from the relevance attention layer, category encoding $\smallh^\smallc_\smallb$, and rating encoding $\smallh^\smallg_\one$ to generate a token $\hat \smally_\smalli$.

\subsection{Training and Inference}
\label{sec:train}

\stitle{Training.}
All the components of the model are trained jointly over $\bign$ training examples in an end-to-end fashion to minimize the negative log-likelihood loss as given below:
\begin{equation*}
\label{eq:model_loss}
 \bigL(\theta)= - \min_\theta \frac{\one}{\bign} \sum_{\smalli=\one}^{\bign} \log p_\theta (\bigy^\smalli | \bigx^\smalli, \bigR^\smalli, \bigC^\smalli, \smallg^\smalli),
\end{equation*}
where $\bigR^\smalli$ represents relevant snippets from app description and reviews retrieved by the IR module. App category and review rating are represented by $\bigC^\smalli$ and $\smallg^\smalli$ respectively, for review-response training example $\smalli$ of form ($\bigx^\smalli$, $\bigy^\smalli$). $\theta$ is a set of trainable parameters of the model. The teacher forcing algorithm that always passes ground truth to the decoder at next time-step, has traditionally been used to achieve faster convergence in training, but it causes an incompatibility in the train and test set-ups. Whereas curriculum learning~\cite{bengio2009curriculum} algorithm passes the current prediction of the decoder to the next time-step to minimize the incompatibility of the train and test set-ups, and enables the model to correct itself, but the model may take longer to train and converge.
In our training, we use a mix of both algorithms with equal probability in the sequence decoder layer to minimize the incompatibility of the train and test set-ups and achieve fast convergence at the same time.

\stitle{Inference.}
We select the model with the best performance on the validation set for inference.
We utilize the beam search algorithm~\cite{wiseman2016sequence} that has been employed in natural language generation tasks like neural machine translation.
While generating the response, it picks multiple alternative tokens (i.e., the ones with high probabilities) from the decoder's probability distribution at every time-step.
The parameter $\bigB$ controls the number of alternatives. At subsequent time-steps, $\bigB$ copies of the decoder are created, each receives a different input from the previous time-step and picks multiple alternative choices. Finally, the output that maximizes the joint probability of the response is selected. While this approach is computationally expensive, it has shown better performance than greedy decoding that picks the word with maximum probability at every time-step. 

\section{Experimental Setup}
\label{sec:experiments}

\begin{table*}[htp]
\footnotesize
\centering
    \caption{
    Sample input reviews along with relevant reviews and app description snippets: relevant reviews and app descriptions not  only  contain  contextual  keywords, but also often provide partial answers to questions raised in input reviews.}
\label{tbl:study_sample}
\begin{tabular}{p{.07\textwidth}|p{.2\textwidth}|p{.42\textwidth}|p{.2\textwidth }}
\hline
\textbf{App} &\textbf{User review} & \textbf{Relevant reviews from the same app} & \textbf{App description} \\\hline \hline
\multirow{3}{.07\textwidth}{Udemy (Education)}&\multirow{3}{.2\textwidth}{... intermittent coverage on my train ride and wanted to watch ... seemingly impossible.}&$\triangleright$ ... \underline{download lecture videos} for offline ... on my train ride to campus ... & \multirow{3}{.2\textwidth}{... \underline{Download courses} to \underline{learn offline}. On the go ...} \\
  & &$\triangleright$ ... problem with video ... It only \underline{work after I downloaded} it. &  \\\hline
\multirow{2}{.07\textwidth}{Uber Eats (Food)}&\multirow{2}{.2\textwidth}{... food was not delivered ... cannot refund the money ...} &$\triangleright$ ... my order was cancelled ... my \underline{money was transferred back} ...& \multirow{2}{.2\textwidth}{Track your food order ...
See the \underline{estimated delivery time }...} \\
 & &$\triangleright$ ... food that is undercooked ... \underline{they refunded} for the entire meal. &  \\\hline
\multirow{2}{.07\textwidth}{Uber (Navigation)}& ... charge cancellation fee ... requires credit or debit card ... &$\triangleright$ ... \underline{do not charge cancellation fees}, trust worthy, \underline{make refunds easily} ...& \multirow{2}{.2\textwidth}{... request a ride ... pay with \underline{credit or cash} in select cities ...} \\\hline
\end{tabular}
\vspace{-0.2cm}
\end{table*}

\subsection{Dataset}
\label{subsec:dataset}
\stitle{Data Collection.}
We crawled $103$ popular apps (those with at least $25$K star ratings and at least $100$ developer replies) across $23$ app categories, and collected over $3.4$ million reviews and more than $570$K review-response pairs.
We collected app name, description, number of star ratings, app category, review text, review time, review rating, developer response, and response time.
Thanks to our app selection criteria, we were able to collect reviews with a much higher response rate: $14.4$\% compared to $2.8$\% in~\cite{hassan2018studying}.

\stitle{Preprocessing.}
User reviews often contain noisy data~\cite{gao2019emerging}. To mitigate this, we performed the following preprocessing steps:
\myNum{i} removed non-English reviews and responses using a language detector~\cite{langdetect}, which ensures a concise and valid vocabulary;
\myNum{ii} performed standard natural language processing (NLP) preprocessing steps such as conversion of letters to lower case, replacement of numbers with ``$<$number$>$'', emails and URLs with ``$<$email$>$'' and ``$<$url$>$'', respectively;
\myNum{iii} replaced greetings and signatures with ``$<$salutation$>$'' and ``$<$signature$>$``, respectively, to preserve user anonymity; and
\myNum{iv} removed reviews and responses with less than four words since such reviews/responses are not likely to contain useful information.
After preprocessing, we obtained $425,618$ review-response pairs and $2,077,674$ reviews with no response.
We found that $47$ out of $103$ apps overlap with RRGen~\cite{rrgen} dataset apps.
Next, we applied the same preprocessing steps to the RRGen dataset, which resulted in $145$K review-response pairs (out of $309$K pairs in the original RRGen dataset). We finally merged the two datasets and the final dataset contains $570,881$ review-response pairs.

We randomly split the review-response pairs of our dataset into training ($530,872$), validation ($19,511$), and test ($19,480$ $\approx 3.5\%$ of dataset) sets.

\stitle{Dataset Analysis.}
Table~\ref{tbl:study_sample} shows a few examples of user reviews, relevant reviews, and app description snippets from our dataset.
The snippets from the relevant reviews and app description contain keywords and similar context that can help synthesize a response.
In fact, the relevant snippets in many cases provide partial answers for the given user review.
For example, while using the Udemy app (see  Table~\ref{tbl:study_sample}), a user was unable to watch a video due to intermittent network coverage on a train ride. In this case, the app description happens to have relevant guidance -- ``download courses to learn offline''.
Luckily, other users who faced a similar issue shared the  solution ``download lecture videos''.

\subsection{Evaluation Metrics}
\label{subsec:evalmetric}
We use quantitative automatic metrics as well as subjective human studies to evaluate the performance of {\oursys} and the competing systems.

\stitle{Automatic Metrics.}
BLEU~\cite{bleupapineni2002} score is a standard automatic metric to evaluate natural language generation solutions such as machine translation~\cite{bahdanau2014neural} and paraphrasing~\cite{siddique2020unsupervised}.
It has been demonstrated to have a positive correlation with human judgements. BLEU-n ($n \in \{1,2,3,4\}$) score $\in [0,100]$ captures the percentage of the n-grams from the synthesized response $\hat \bigy$ that also co-occur in the the ground truth $\bigy$, where $0$ means no matching n-grams and $100$ means a perfect match.
We utilize BLEU-4, which is considered a standard metric. Moreover, we also use the recall-based automatic metric  ROUGE~\cite{lin-2004-rouge}, which measures n-grams overlap between the synthesized response $\hat \bigy$ and the ground truth $\bigy$. We use ROUGE-L that identifies the longest co-occurrence of n-grams using the Longest Common Subsequence (LCS)~\cite{lin-och-2004-automatic}, which naturally evaluates sentence structure similarity.

\stitle{Human Study.}
We conducted a subjective human study to evaluate the quality of the responses generated by {\oursys} and other competing methods with respect to the developer responses.
We made use of Amazon Mechanical Turk,
a crowd sourcing platform for human evaluation, where human evaluators rate the quality of the response on a scale of $1-5$, $1$ being the worst and $5$ being the best. We asked human evaluators to consider three aspects in their evaluations:
\myNum{i} the response is specific to the app; and
\myNum{ii} whether the response addresses the concern of the user raised in the input review;
\myNum{iii} the language fluency and the grammatical correctness of the response.
We randomly selected $150$ generated responses by {\oursys} and the  competing systems for the same reviews, and each response was scored by five different human evaluators who are familiar with the respective app.  

\subsection{Competing Approaches}
\label{subsec:competing}
We compare {\oursys} with two IR baselines, an MRC model R-Net~\cite{wang2017r}, and the state-of-the-art response generation system RRGen~\cite{rrgen}. We briefly describe each competing method below: 

\begin{description}[leftmargin=1.2\parindent,labelindent=-5pt, itemsep=-0pt]
\item \textbf{IR-Reviews}:
This baseline builds an index for the reviews in the training set.
For an input review and app name from the test set, it retrieves the most relevant indexed review (i.e., top-$1$) as a response.

\item \textbf{IR-Response}:
This baseline builds an index for the developer responses in the training set.
For an input review and app name from the test set, it retrieves the most relevant indexed developer response  (i.e., top-$1$) as a response.

\item \textbf{R-Net}~\cite{wang2017r}:
This system was proposed for machine reading comprehension style question answering, and it achieves state-of-the-art results on SQuAD~\cite{rajpurkar2016squad} and MS-MARCO~\cite{nguyen2016ms} datasets.
MRC models require annotated answer spans in a passage (app description and reviews in the context of response synthesis) for supervision, which are not available in our dataset, and acquiring such manual annotations is laborious.
We annotated spans using BLEU-2 heuristics: the reviews that achieve the maximum BLEU-2 score with the developer response are considered a span.
R-Net obtains question matching representation of a passage by passing the question and the passage through a GRU, then self-matching attention mechanism is employed to refine the generated repreentation.

\item \textbf{RRGen}~\cite{rrgen}:
RRGen uses attention-based seq2seq for producing a response for an input review. It conditions the response generation on app category, rating, review length, review sentiment, and a set of dictionary-based keywords.
RRGen achieves current state-of-the-art results on BLEU~\cite{bleupapineni2002} metric for the response generation task. Since RRGen outperformed NNGen~\cite{liu2018neural} and NMT~\cite{bahdanau2014neural} models, we do not consider them as competing approaches in this work.  We used RRGen's open-source implementation ~\cite{rrgen-github} in our experiments.
\end{description}

\subsection{Implementation Details}
\label{subsec:impl}
We used Apache Lucene~\cite{lucene} and the BM25~\cite{bm25} scoring algorithm in our IR module, which retrieves $4$ snippets from the relevant user reviews for an app and $1$ snippet from the app description. We implemented {\oursys} in PyTorch~\cite{ketkar2017introduction}. The word embedding dimensions were set to $128$ and the vocabulary size was set to $10,000$.
Based on the word length outlier analysis, we set the maximum length for the review, snippets, app category, rating, and response to $75$, $50$, $4$, $1$, and $120$, respectively.
{\oursys} uses $128$ LSTM units as an RNN that has $2$ layers for all the encoders (review, snippets, app rating, and category encoders) and the decoder.
We set the batch size to $128$ and the dropout rate to $0.2$ (i.e., to avoid over-fitting to training set).
We trained the neural network for $25$ epochs using Adam Optimizer and employ negative log likelihood loss with a learning rate of $0.01$.

\section{Results}
\label{sec:evals}

\begin{table}[t]
\centering
\caption{Automatic metrics results: p$_n$ represents $n$-gram precision for the ground truth and synthesized response.}
\label{tbl:quant_results}
\begin{tabular}{lrrrrrr} \hline
Method & BLEU & ROUGE-L & p$_1$ & p$_2$ & p$_3$ & p$_4$  \\ \hline
IR Reviews & 13.67 & 12.76 & 21.49 & 15.56 & 11.44 & 8.56 \\
IR Response & 19.19 & 17.99 & 27.18 & 20.89 & 17.14 & 14.88 \\
R-Net & 29.16 & 39.92 & 42.89 & 29.83 & 23.19 & 16.72\\
RRGen & 34.55 & 46.26 & 50.38 & 37.54 & 28.25 & 22.63 \\ \hline
{\oursys} & \textbf{42.22} & \textbf{51.89} & \textbf{56.99} & \textbf{44.50} & \textbf{36.56} & \textbf{30.83} \\
\hline
\end{tabular}
\end{table}

\subsection{Automatic Metrics}
\label{subsec:autoevals}

\stitle{Performance comparison.}
Table~\ref{tbl:quant_results} shows the BLEU, ROUGE-L, and n-gram precision scores for all competing systems.
As expected, the IR baselines IR Reviews and IR Response produce the worst results ($13.67$ and $19.19$ BLEU scores, respectively), since they can not generate a response and are only capable of retrieving the most relevant reviews or responses from the training set.
The MRC model R-Net produces mediocre results, mainly because of the limitation of MRC models that they attempt to predict spans from the related documents (i.e., app description and reviews), which rarely contain perfect responses.
However, R-Nets's BLEU score of $29.16$ shows that snippets from the app description and reviews can be helpful for synthesizing a high-quality response.
{\oursys} achieves the best results on all automatic metrics; specifically, it outperforms the state-of-the-art system RRGen by $22.20\%$ on BLEU and $12.17\%$ on ROUGE-L. The outstanding performance of {\oursys} is attributed to the core idea of this paper, i.e., fusing MRC with attention-based seq2seq.
The MRC model, coupled with our IR module, discovers relevant app-specific knowledge from relevant reviews and app descriptions, and produces a review-aware representation that is associated with the input review and its important keywords. The seq2seq model learns from the available review-response training pairs. The seq2seq decoder fuses knowledge from both MRC and seq2seq encoders, and thus learns to synthesize responses that are not only fluent and relevant but also app-aware.

\begin{table}[t]
\centering
\caption{Contribution of each component on {\oursys}'s performance. The attention-based seq2seq is the basic component of {\oursys}; thus, it is present in all configurations.}
\label{tbl:ablation}
\begin{tabular}{lcc} \hline
Component & BLEU & ROUGE-L  \\ \hline
Attentional seq2seq & 21.36 & 26.19  \\ \hline
+ Rating & 24.20 & 31.80 \\
+ Category & 30.97 & 44.57  \\
+ Rating + Category & 31.81 & 44.96  \\
+ Description & 34.75 & 49.71  \\
+ Rating + Category + Description & 36.11 & 49.98  \\
+ Description + Review & 38.31 & 50.39  \\
 \hline
{\oursys} (Rating + Category + Desc. + Review) & \textbf{42.22} & \textbf{51.89}  \\
\hline
\end{tabular}
\vspace{-0.4cm}
\end{table}

\stitle{Ablation study.}
Table~\ref{tbl:ablation} shows the effect of each component on the performance of {\oursys}; that is, the table shows the performance of {\oursys} when certain components are enabled.
When only the attention-based seq2seq is enabled, {\oursys} achieves a $21.36$ and $26.19$ BLEU and ROUGE-L scores, respectively.
Enabling the use of app features such as review rating and app category results in a modest improvement of $+2.84$ and $+9.61$ on BLEU score, respectively. While other features such as review length and sentiment may, in principle, improve the performance of the system, it has been confirmed in~\cite{rrgen} that such features merely cause small improvements.
We highlight that enabling the use of our fused architecture (configuration Att. seq2seq + Description + Review) produces the maximum gain on all the metrics: $+16.95$ on BLEU and $+24.2$ on ROUGE-L metrics. 
Moreover, if no relevant reviews are available for a given app (configuration Att. seq2seq + Rating + Category + Description), {\oursys} continues to provide better performance compared to the state-of-the-art system RRGen ($36.11$ vs. $34.55$  BLEU scores).  Several other configurations are also provided in Table~\ref{tbl:ablation} for completeness.

\stitle{Effect of hyperparameters.}
Table~\ref{tbl:parameters} shows the effect of adjusting the hyperparameters on the overall performance of {\oursys}.
Configuration 5, which has 128 hidden and word embedding dimensions and uses up to top-5 relevant snippets, outperforms the other configurations.
Increasing the size of the hidden units, word embedding dimensions, or  the number of relevant snippets does not necessarily improve the performance.
Additionally, configuration 3 and 5 are similar with one exception: the number of layers.
Increasing the number of layers improves the results since more layers can capture better representations.
Note that even configuration 1, which produces the worst results among the reported configurations, outperforms the state-of-the-art system RRGen.
This confirms that our performance is not due to parameter tuning, but rather due to fusing seq2seq and MRC architectures.

\begin{table}[t]
\centering
\caption{Effects of different hyperparameters on the performance of {\oursys}. $L$ is the number of layers, $H$ is the number hidden dimensions, $E$ is the number of embedding dimensions, and $S$ is the number of relevant snippets.}
\label{tbl:parameters}
\begin{tabular}{lcrrrrr} \hline
Sr. \# & Configuration & BLEU & ROUGE-L  \\ \hline
1 & $L$=2, $H$=256, $E$ = 256, $S$ = 5 & 35.95 & 46.16  \\ 
2 & $L$=2, $H$=128, $E$ = 256, $S$ = 5 & 36.80 & 49.30  \\
3 & $L$=1, $H$=128, $E$ = 128, $S$ = 5 & 39.31 & 49.69 \\
4 & $L$=2, $H$=128, $E$ = 128, $S$ = 10 & 41.72 & 51.58 \\
5 & $L$=2, $H$=128, $E$ = 128, $S$ = 5 & \textbf{42.22} & \textbf{51.89}  \\
\hline
\end{tabular}
\end{table}

\subsection{Human Study}
\label{subsec:humanstudy}
We conducted a human study to evaluate the quality of the responses generated by each of the competing systems from a human perspective.
We limit this study to the best performing systems based on the results of the automatic metric evaluation. Specifically, we compare {\oursys}, R-Net, and RRGen in this study. We randomly selected $150$ input reviews and generated their respective responses using each of the competing systems.
For each input review, we randomly selected 5 human evaluators to score the respective responses without knowing which responses were generated by which method or whether the response was produced by a developer. Additionally, we required that the evaluators are familiar with the app whose responses are under investigation;
in total, $750$ evaluators participated in the evaluations.
Evaluators were asked to judge a response based on three criteria: app-specificity, whether it contains information that addresses user concerns, and fluency.
We report average scores with confidence intervals for responses generated by developers (our ground truth) and each of the competing systems in Table~\ref{tbl:human}. 
The responses generated by {\oursys} achieve the best scores on all the aspects when compared to other generative methods. The statistical significance test result ($p-value$ $<$ 0.01)~\cite{woolson2007wilcoxon} also shows that the responses generated by our method achieve statistically significant improvement on all three criteria when compared with R-Net and RRGen. Moreover, responses generated by {\oursys} achieve very close scores to those manually generated by developers. While {\oursys} produces app-specific high-quality responses, these are often not as fluent as manually written responses. This is a common limitation in most conditional natural language generation tasks. In future, we plan to improve the fluency of {\oursys}'s responses by incorporating a neural language model.
Overall, the results of our human study show that the responses generated by {\oursys} are not only better than other automatic response generation methods, but also on a par with the developers' manually generated responses.

\subsection{Sample Responses}
\label{subsec:discussion}
We present in Table~\ref{tbl:samples} sample responses generated by app developers, R-Net, RRGen, and {\oursys}. In what follows, we highlight several interesting examples. Consider the UC Browser app which offers a download option. When a user faces problems while using this option, {\oursys} is able to discover the resume feature in the download option through relevant snippets, and it offers a possible solution for the problem by including ``try resume again it starts downloading...'' in the response. Similarly, {\oursys} learns whether ``clear the cache of the app'' can possibly solve the issue for the Adobe Reader app (see Table~\ref{tbl:samples}), and forwards this information by assigning high weights for the important keywords to the sequence decoder to generate app-aware responses. This example highlights how {\oursys} generates app-specific responses.
{\oursys} takes advantage of the ever increasing relevant snippets from other reviews of the same app, which are likely to have answers to other reviews. This enables {\oursys} to generate responses that directly address user concerns as opposed to responses that are vaguely related to the input review. Consider the input review ``that is not free. they told free but not'', where the user is concerned about the app content not being free. {\oursys} is able to learn that the user concern is about pricing and it communicates a possible remedy by responding with ``... you can use free course filter...'' (see Table~\ref{tbl:samples}).

\begin{table}[t!]
\centering
\caption{Human study results: {\oursys} outperforms all competing systems, and the quality of its responses is close to that of expertly written responses (Developer responses).}
\label{tbl:human}
\begin{tabular}{l|r|r|r} \hline
Method & App-specific & Addresses Concern & Fluent   \\ \hline\hline
R-Net & 2.911$\pm$0.041 & 2.881$\pm$0.042 & 2.856$\pm$0.043  \\
RRGen & 2.915$\pm$0.041 &	2.769$\pm$0.044 & 2.957$\pm$0.042  \\ 
\textsc{AARSynth} & \textbf{3.408}$\pm$0.038 &	\textbf{3.403}$\pm$0.038 &	\textbf{3.456}$\pm$0.038   \\\hline\hline
Developer & 3.436$\pm$0.039 & 3.420$\pm$0.039 & 3.575$\pm$0.037   \\ \hline
\end{tabular}
\end{table}

\begin{table*}[t!]
\footnotesize
\caption{Sample user reviews, developer responses and generated responses using R-Net, RRGen and \textsc{AARSynth}.}
\label{tbl:samples}
\begin{tabular}{p{.14\textwidth}|p{.08\textwidth}| p{.72\textwidth}} \hline
\textbf{Review} & \textbf{Method} &  \textbf{Response}\\ \hline \hline
\multirow{7}{.14\textwidth}{that is not free. they told free but not} & \multirow{2}{.08\textwidth}{Developer} & $<$salutation$>$ i'm sorry for the inconvenience you are having with the price on udemy. could you please contact us via $<$email$>$ so we can assist you better.\\
& \multirow{2}{.08\textwidth}{R-Net} & $<$salutation$>$ it certainly sounds frustrating and unpleasant experience. we can look into this for you. let us know by sending a quick note to $<$url$>$.\\
& RRGen & $<$salutation$>$ please send a note us at: $<$email$>$. \\ 
& \multirow{2}{.08\textwidth}{{\oursys}} & $<$salutation$>$ udemy is free, but all online contents are not free. \underline{you can use free course filter}. please contact us via $<$email$>$ for further assistance. \\ \hline
\multirow{7}{.14\textwidth}{not able to share selection? you must be joking. this is the de facto standard reader and behaves in such an autistic way.} & 
Developer &  please update your "adobe reader app". issue has been fixed. $<$signature$>$ \\ 
& \multirow{2}{.08\textwidth}{R-Net} &  $<$salutation$>$ we are sorry to hear about your experience with about request you to reach out to us at $<$url$>$ and we'll will take into this.\\
& \multirow{2}{.08\textwidth}{RRGen} &  $<$salutation$>$ we have facing difficulties with our app. email a concerns, please contact free to contact with support team at $<$email$>$.\\
& \multirow{2}{.08\textwidth}{{\oursys}} &  $<$salutation$>$ we are sorry to hear about your experience. \underline{please clear the cache of app from the phone settings}. if you are still unable to make it work, reach out to us at $<$email$>$ and we'll will look into this. \\ \hline
\multirow{6}{.14\textwidth}{fare must be deduce.} & \multirow{2}{.08\textwidth}{Developer} & 
$<$salutation$>$ we appreciate your feedback and be work towards make this better. if there's anything else we can take a look into for you please let us know to $<$email$>$ so we can further connect.\\ 
& R-Net &  we like to hear into this. please email us at $<$email$>$.\\
& RRGen &  we appreciate your concern and would like to investigate into your issue. please email us directly afterwards $<$email$>$ and $<$number$>$.\\
& \multirow{2}{.08\textwidth}{{\oursys}} &  $<$salutation$>$ we love to hear your feedback. we believe that \underline{fare must be higher but comfortable rides than others}. Our team is working on this aspect as well. please contact us directly : $<$email$>$. \\ \hline
\multirow{7}{.14\textwidth}{movie downloading stuck at $<$number$>$ - $<$number$>$. worst app.} & \multirow{3}{.08\textwidth}{Developer} & $<$salutation$>$ we regret the inconvenience you have experienced. pls send us more details via menu-feedback page(open in uc browser).you may clear cache, change download path, turn on/off cloud boost, change access point and then $<$url$>$ a nice day dear! \\
& R-Net &  $<$salutation$>$ we feel sorry for the inconvenience. if you resume again it starts downloading.\\
& RRGen &  $<$salutation$>$ we are sorry to hear that. if you will contact to $<$email$>$ with the version from google app.\\
& \multirow{2}{.08\textwidth}{{\oursys}} & $<$salutation$>$ we regret the inconvenience you have experienced. \underline{if you try resume again it starts downloading from} $<$number$>$. if issue persists, we suggest you to visit $<$url$>$.\\ \hline
\end{tabular}
\end{table*}

\section{Related Work}
\label{sec:related}

\stitle{Dialogue Response Generation.}
Response generation is one of the prominent topics in the field of NLP where a system attempts to provide an answer to a given question~\cite{ji2014information,zeng2019you}. IR systems~\cite{manning2008introduction, salton1975vector, chen2011evaluating} aim to retrieve the best matching response, and machine translation based approaches~\cite{koehn2007moses, ritter2011data, vinyals2015neural} try to generate the response for a given question.
Touch your heart~\cite{tianran2018touch} proposes to generate tone-aware responses for customer care on social media.
RRGen~\cite{rrgen} employs features such as category, review sentiments and so on, to generate the responses for user reviews on Google Play.
Since RRGen is trained on the review-response pairs of many apps, it can only provide generic responses, and it struggles to answer app-specific reviews.
Recently, Google Play also started offering suggested replies~\cite{googleplayresponse}. This system also suggests generic replies as others. 
We utilize snippets from app description and reviews (retrieved through IR module) along with other features to synthesize a fluent and app-aware response that addresses user concerns raised in an input review by fusing seq2seq with MRC models that employ attention mechanisms.

\stitle{Question Answering.} 
Question answering is an extensively studied research problem, which includes several sub-tasks and datasets. The selection of answers in community question answering~\cite{wu2018question} and ranking question-answer pairs~\cite{yoon2018learning} are related to response synthesis; however, in response synthesis, the goal is to generate coherent and fluent answers rather than answer selection.
Similarly, open-domain question answering by reasoning over knowledge bases~\cite{yin2016simple} and large open-domain sources such as Wikipedia~\cite{tay2018densely} can not be employed for the task of review response generation, as knowledge bases are not helpful for synthesizing responses for reviews. 
Last but not least, closed-domain question answering datasets such as SQuAD~\cite{rajpurkar2016squad}, SearchQA~\cite{dunn2017searchqa}, and MS MARCO~\cite{nguyen2016ms} assume that the answer to the question is a span from the accompanying set of document(s), whereas developer response is rarely a span from the related reviews or application description. 

\stitle{Analysis of App Reviews.}
Many studies collected and examined different aspects of app reviews.
ChangeAdvisor~\cite{palomba2017recommending} uses app reviews to extract useful feedback to recommend software maintenance changes.
The authors in~\cite{grano2018exploring} studied 
user reviews to improve automated testing.
The authors 
in~\cite{gao2019emerging} used app reviews to detect real-time emerging issues in the apps, others used GitHub to study issues~\cite{farooq2018, farooq2020}. 
The authors in~\cite{maalej2015bug} used probabilistic techniques to classify reviews into bug report, feature requests and so on.
Sentiment analysis on app features is performed in ~\cite{mike2010sentiment, guzman2014users}.
Similarly, \textsc{Mara}~\cite{iacob2013retrieving} uses reviews to predict app feature requests. 
\cite{hassan2018studying} analyzed $4.5$ million reviews and highlighted the importance of developer replies to app reviews. Our research extends the findings of these works by collecting a large dataset of reviews and using the related review snippets to generate app-aware responses.  
\section{Conclusion}
\label{sec:conclusion}
We have presented {\oursys}, a system for automatically synthesizing app-specific responses to mobile app reviews. 
{\oursys} is motivated by the impossibility of manually responding to the large, growing number of reviews.
{\oursys} enables developers to enjoy the benefits of providing timely and relevant responses including improved ratings and wider adoption of their apps.
Our experimental evaluation of {\oursys} using a large corpus of reviews and responses from Google Play showed that it significantly outperforms the existing approaches: an improvement of 22.2\%  in BLEU-4 score over the state-of-the-art system.
Moreover, human evaluators' ratings of the responses generated by {\oursys} and other competing methods suggest that, with a statistically significant difference, {\oursys}'s responses are app-specific, better address the concerns raised in an input review, and are more linguistically fluent. 
The main novelty of {\oursys} is augmenting seq2seq models with app-specific information, which is made possible due to utilizing information retrieval techniques and our fusion of seq2seq and machine reading comprehension neural architectures.

\section*{Acknowledgment}
\noindent This work is supported in part by the National Science Foundation under grants IIS-1838222 and IIS-1901379.

\bibliographystyle{IEEEtran}
\bibliography{sample-base.bib}

\begin{thebibliography}{10}
\providecommand{\url}[1]{#1}
\csname url@samestyle\endcsname
\providecommand{\newblock}{\relax}
\providecommand{\bibinfo}[2]{#2}
\providecommand{\BIBentrySTDinterwordspacing}{\spaceskip=0pt\relax}
\providecommand{\BIBentryALTinterwordstretchfactor}{4}
\providecommand{\BIBentryALTinterwordspacing}{\spaceskip=\fontdimen2\font plus
\BIBentryALTinterwordstretchfactor\fontdimen3\font minus
  \fontdimen4\font\relax}
\providecommand{\BIBforeignlanguage}[2]{{%
\expandafter\ifx\csname l@#1\endcsname\relax
\typeout{** WARNING: IEEEtran.bst: No hyphenation pattern has been}%
\typeout{** loaded for the language `#1'. Using the pattern for}%
\typeout{** the default language instead.}%
\else
\language=\csname l@#1\endcsname
\fi
#2}}
\providecommand{\BIBdecl}{\relax}
\BIBdecl

\bibitem{statista}
S.~O\'Dea, ``Number of smartphone users worldwide from 2016 to 2021,''
  \url{https://www.statista.com/statistics/330695/number-of-smartphone-users-worldwide/},
  accessed: 2020-04-28.

\bibitem{googleplayresponse}
K.~Glick, ``Making it easier to respond to and improve user reviews,''
  \url{https://android-developers.googleblog.com/2019/05/whats-new-in-play.html},
  2020, accessed: 2020-04-28.

\bibitem{rrgen}
C.~{Gao}, J.~{Zeng}, X.~{Xia}, D.~{Lo}, M.~R. {Lyu}, and I.~{King},
  ``Automating app review response generation,'' in \emph{34th IEEE/ACM
  International Conference on Automated Software Engineering}, 2019, pp.
  163--175.

\bibitem{sutskever2014sequence}
I.~Sutskever, O.~Vinyals, and Q.~V. Le, ``Sequence to sequence learning with
  neural networks,'' in \emph{Advances in neural information processing
  systems}, 2014, pp. 3104--3112.

\bibitem{wang2017r}
W.~Wang, N.~Yang, F.~Wei, B.~Chang, and M.~Zhou, ``R-net: Machine reading
  comprehension with self-matching networks,'' \emph{Microsoft Research Asia,
  Beijing, China, Tech. Rep}, vol.~5, 2017.

\bibitem{seo2016bidirectional}
M.~Seo, A.~Kembhavi, A.~Farhadi, and H.~Hajishirzi, ``Bidirectional attention
  flow for machine comprehension,'' \emph{arXiv preprint:1611.01603}.

\bibitem{hochreiter1997lstm}
S.~Hochreiter and J.~Schmidhuber, ``Lstm can solve hard long time lag
  problems,'' in \emph{Advances in neural information processing systems},
  1997, pp. 473--479.

\bibitem{vaswani2017attention}
A.~Vaswani, N.~Shazeer, N.~Parmar, J.~Uszkoreit, L.~Jones, A.~N. Gomez,
  {\L}.~Kaiser, and I.~Polosukhin, ``Attention is all you need,'' in
  \emph{Advances in neural information processing systems}, 2017, pp.
  5998--6008.

\bibitem{lucene}
A.~Lucene, ``Apache lucene,'' \url{https://lucene.apache.org/}.

\bibitem{pugh1990skip}
W.~Pugh, ``Skip lists: a probabilistic alternative to balanced trees,''
  \emph{Communications of the ACM}, vol.~33, no.~6, pp. 668--676, 1990.

\bibitem{baeza1999modern}
R.~Baeza-Yates and B.~Ribeiro-Neto, ``Modern information retrieval
  addison-wesley longman,'' \emph{Reading MA}, 1999.

\bibitem{bm25}
S.~E. Robertson and K.~S. Jones, ``Relevance weighting of search terms,''
  \emph{Journal of the American Society for Information science}, vol.~27,
  no.~3, pp. 129--146, 1976.

\bibitem{manning2014stanford}
C.~D. Manning, M.~Surdeanu, J.~Bauer, J.~R. Finkel, S.~Bethard, and
  D.~McClosky, ``The stanford corenlp natural language processing toolkit,'' in
  \emph{Proceedings of 52nd annual meeting of the association for computational
  linguistics: system demonstrations}, 2014, pp. 55--60.

\bibitem{bengio2009curriculum}
Y.~Bengio, J.~Louradour, R.~Collobert, and J.~Weston, ``Curriculum learning,''
  in \emph{Proceedings of the 26th annual international conference on machine
  learning}, 2009, pp. 41--48.

\bibitem{wiseman2016sequence}
S.~Wiseman and A.~M. Rush, ``Sequence-to-sequence learning as beam-search
  optimization,'' \emph{arXiv preprint:1606.02960}, 2016.

\bibitem{hassan2018studying}
S.~Hassan, C.~Tantithamthavorn, C.-P. Bezemer, and A.~E. Hassan, ``Studying the
  dialogue between users and developers of free apps in the google play
  store,'' \emph{Empirical Software Engineering}, vol.~23, no.~3, pp.
  1275--1312, 2018.

\bibitem{gao2019emerging}
C.~Gao, W.~Zheng, Y.~Deng, D.~Lo, J.~Zeng, M.~R. Lyu, and I.~King, ``Emerging
  app issue identification from user feedback: experience on wechat,'' in
  \emph{IEEE/ACM 41st International Conference on Software Engineering:
  Software Engineering in Practice}, 2019, pp. 279--288.

\bibitem{langdetect}
M.~Danilák, ``Port of google\'s language-detection library to python,''
  \url{https://github.com/Mimino666/langdetect}, 2020, accessed: 2020-04-28.

\bibitem{bleupapineni2002}
K.~Papineni, S.~Roukos, T.~Ward, and W.-J. Zhu, ``Bleu: a method for automatic
  evaluation of machine translation,'' in \emph{Proceedings of the 40th annual
  meeting on association for computational linguistics}.\hskip 1em plus 0.5em
  minus 0.4em\relax Association for Computational Linguistics, 2002, pp.
  311--318.

\bibitem{bahdanau2014neural}
D.~Bahdanau, K.~Cho, and Y.~Bengio, ``Neural machine translation by jointly
  learning to align and translate,'' \emph{arXiv preprint:1409.0473}, 2014.

\bibitem{siddique2020unsupervised}
A.~Siddique, S.~Oymak, and V.~Hristidis, ``Unsupervised paraphrasing via deep
  reinforcement learning,'' in \emph{Proceedings of the 26th ACM SIGKDD
  International Conference on Knowledge Discovery \& Data Mining}, 2020, pp.
  1800--1809.

\bibitem{lin-2004-rouge}
C.-Y. Lin, ``{ROUGE}: A package for automatic evaluation of summaries,'' in
  \emph{Text Summarization Branches Out}.\hskip 1em plus 0.5em minus
  0.4em\relax Barcelona, Spain: Association for Computational Linguistics, Jul.
  2004, pp. 74--81.

\bibitem{lin-och-2004-automatic}
C.-Y. Lin and F.~J. Och, ``Automatic evaluation of machine translation quality
  using longest common subsequence and skip-bigram statistics,'' in
  \emph{Proceedings of the 42nd Annual Meeting of the Association for
  Computational Linguistics}, Barcelona, Spain, Jul. 2004, pp. 605--612.

\bibitem{rajpurkar2016squad}
P.~Rajpurkar, J.~Zhang, K.~Lopyrev, and P.~Liang, ``Squad: 100,000+ questions
  for machine comprehension of text,'' \emph{arXiv preprint:1606.05250}, 2016.

\bibitem{nguyen2016ms}
T.~Nguyen, M.~Rosenberg, X.~Song, J.~Gao, S.~Tiwary, R.~Majumder, and L.~Deng,
  ``Ms marco: a human-generated machine reading comprehension dataset,'' 2016.

\bibitem{liu2018neural}
Z.~Liu, X.~Xia, A.~E. Hassan, D.~Lo, Z.~Xing, and X.~Wang,
  ``Neural-machine-translation-based commit message generation: how far are
  we?'' in \emph{Proceedings of the 33rd ACM/IEEE International Conference on
  Automated Software Engineering}, 2018, pp. 373--384.

\bibitem{rrgen-github}
C.~Gao, ``Repository for the rrgen,'' \url{https://github.com/armor-ai/RRGen},
  2020, accessed: 2020-04-28.

\bibitem{ketkar2017introduction}
N.~Ketkar, ``Introduction to pytorch,'' in \emph{Deep learning with
  python}.\hskip 1em plus 0.5em minus 0.4em\relax Springer, 2017, pp. 195--208.

\bibitem{woolson2007wilcoxon}
R.~Woolson, ``Wilcoxon signed-rank test,'' \emph{Wiley encyclopedia of clinical
  trials}, pp. 1--3, 2007.

\bibitem{ji2014information}
Z.~Ji, Z.~Lu, and H.~Li, ``An information retrieval approach to short text
  conversation,'' \emph{arXiv preprint:1408.6988}, 2014.

\bibitem{zeng2019you}
J.~Zeng, J.~Li, Y.~He, C.~Gao, M.~R. Lyu, and I.~King, ``What you say and how
  you say it: Joint modeling of topics and discourse in microblog
  conversations,'' \emph{Transactions of the Association for Computational
  Linguistics}, vol.~7, pp. 267--281, 2019.

\bibitem{manning2008introduction}
C.~D. Manning, P.~Raghavan, and H.~Sch{\"u}tze, \emph{Introduction to
  information retrieval}.\hskip 1em plus 0.5em minus 0.4em\relax Cambridge
  university press, 2008.

\bibitem{salton1975vector}
G.~Salton, A.~Wong, and C.-S. Yang, ``A vector space model for automatic
  indexing,'' \emph{Communications of the ACM}, vol.~18, no.~11, pp. 613--620,
  1975.

\bibitem{chen2011evaluating}
G.~Chen, E.~Tosch, R.~Artstein, A.~Leuski, and D.~Traum, ``Evaluating
  conversational characters created through question generation,'' in
  \emph{Twenty-Fourth International FLAIRS Conference}, 2011.

\bibitem{koehn2007moses}
P.~Koehn, H.~Hoang, A.~Birch, C.~Callison-Burch, M.~Federico, N.~Bertoldi,
  B.~Cowan, W.~Shen, C.~Moran, R.~Zens \emph{et~al.}, ``Moses: Open source
  toolkit for statistical machine translation,'' in \emph{Proceedings of the
  45th annual meeting of the association for computational linguistics
  companion}, 2007, pp. 177--180.

\bibitem{ritter2011data}
A.~Ritter, C.~Cherry, and W.~B. Dolan, ``Data-driven response generation in
  social media,'' in \emph{Proceedings of the conference on empirical methods
  in natural language processing}.\hskip 1em plus 0.5em minus 0.4em\relax
  Association for Computational Linguistics, 2011, pp. 583--593.

\bibitem{vinyals2015neural}
O.~Vinyals and Q.~Le, ``A neural conversational model,'' \emph{arXiv
  preprint:1506.05869}, 2015.

\bibitem{tianran2018touch}
T.~Hu, A.~Xu, Z.~Liu, Q.~You, Y.~Guo, V.~Sinha, J.~Luo, and R.~Akkiraju,
  ``Touch your heart: A tone-aware chatbot for customer care on social media,''
  in \emph{Proceedings of the 2018 CHI Conference on Human Factors in Computing
  Systems}, ser. CHI ’18.\hskip 1em plus 0.5em minus 0.4em\relax ACM, 2018.

\bibitem{wu2018question}
W.~Wu, X.~Sun, and H.~Wang, ``Question condensing networks for answer selection
  in community question answering,'' in \emph{Proceedings of the 56th Annual
  Meeting of the Association for Computational Linguistics (Volume 1: Long
  Papers)}, 2018, pp. 1746--1755.

\bibitem{yoon2018learning}
S.~Yoon, J.~Shin, and K.~Jung, ``Learning to rank question-answer pairs using
  hierarchical recurrent encoder with latent topic clustering,'' in
  \emph{Proceedings of the 2018 Conference of the North American Chapter of the
  Association for Computational Linguistics: Human Language Technologies,
  Volume 1 (Long Papers)}, 2018, pp. 1575--1584.

\bibitem{yin2016simple}
W.~Yin, M.~Yu, B.~Xiang, B.~Zhou, and H.~Sch{\"u}tze, ``Simple question
  answering by attentive convolutional neural network,'' \emph{arXiv
  preprint:1606.03391}, 2016.

\bibitem{tay2018densely}
Y.~Tay, A.~T. Luu, S.~C. Hui, and J.~Su, ``Densely connected attention
  propagation for reading comprehension,'' in \emph{Advances in Neural
  Information Processing Systems}, 2018, pp. 4906--4917.

\bibitem{dunn2017searchqa}
M.~Dunn, L.~Sagun, M.~Higgins, V.~U. Guney, V.~Cirik, and K.~Cho, ``Searchqa: A
  new q\&a dataset augmented with context from a search engine,'' \emph{arXiv
  preprint:1704.05179}, 2017.

\bibitem{palomba2017recommending}
F.~Palomba, P.~Salza, A.~Ciurumelea, S.~Panichella, H.~Gall, F.~Ferrucci, and
  A.~De~Lucia, ``Recommending and localizing change requests for mobile apps
  based on user reviews,'' in \emph{Proceedings of the 39th International
  Conference on Software Engineering}, ser. ICSE ’17.\hskip 1em plus 0.5em
  minus 0.4em\relax IEEE Press, 2017, p. 106–117.

\bibitem{grano2018exploring}
G.~Grano, A.~Ciurumelea, S.~Panichella, F.~Palomba, and H.~C. Gall, ``Exploring
  the integration of user feedback in automated testing of android
  applications,'' in \emph{IEEE 25th International Conference on Software
  Analysis, Evolution and Reengineering}.\hskip 1em plus 0.5em minus
  0.4em\relax IEEE, 2018, pp. 72--83.

\bibitem{farooq2018}
U.~Farooq and Z.~Zhao, ``Runtimedroid: Restarting-free runtime change handling
  for android apps,'' in \emph{Proceedings of the 16th Annual International
  Conference on Mobile Systems, Applications, and Services}, 2018.

\bibitem{farooq2020}
U.~Farooq, Z.~Zhao, M.~Sridharan, and I.~Neamtiu, ``Livedroid: Identifying and
  preserving mobile app state in volatile runtime environments,'' in
  \emph{Proceedings of the ACM on Programming Languages (PACMPL), Volume 4,
  Article 160, Issue OOPSLA, 2020}.\hskip 1em plus 0.5em minus 0.4em\relax ACM,
  2020.

\bibitem{maalej2015bug}
W.~Maalej and H.~Nabil, ``Bug report, feature request, or simply praise? on
  automatically classifying app reviews,'' in \emph{IEEE 23rd international
  requirements engineering conference}.\hskip 1em plus 0.5em minus 0.4em\relax
  IEEE, 2015, pp. 116--125.

\bibitem{mike2010sentiment}
T.~Mike, B.~Kevan, P.~Georgios, and C.~Di, ``Sentiment in short strength
  detection informal text,'' \emph{JASIST}, vol.~61, no.~12, pp. 2544--2558,
  2010.

\bibitem{guzman2014users}
E.~Guzman and W.~Maalej, ``How do users like this feature? a fine grained
  sentiment analysis of app reviews,'' in \emph{IEEE 22nd international
  requirements engineering conference}.\hskip 1em plus 0.5em minus 0.4em\relax
  IEEE, 2014, pp. 153--162.

\bibitem{iacob2013retrieving}
C.~Iacob and R.~Harrison, ``Retrieving and analyzing mobile apps feature
  requests from online reviews,'' in \emph{2013 10th working conference on
  mining software repositories}.\hskip 1em plus 0.5em minus 0.4em\relax IEEE,
  2013, pp. 41--44.

\end{thebibliography}

\balance

\end{document}